# Adding variances of independent probability distributions to estimate probabilities of replication

Huw Llewelyn[1,*]

[1]Department of Mathematics, Aberystwyth University, Penglais, SY23 3BZ, Ceredigion, United Kingdom

[*]Department of Mathematics, Aberystwyth University, Penglais, Aberystwyth, SY23 3B, United Kingdom email-hul2@aber.ac.uk

## Abstract

If the prior probability distributions of all possible hypothetical true means and all possible observed means of a continuous variable are conditional on the universal set of all numbers (i.e., before the nature of a study is known and a Bayesian prior distribution can be estimated), that prior probability distribution will be uniform. It would follow that a Gaussian probability distribution and a Gaussian likelihood distribution based on the same data set are identical. Replication involves doing two independent studies and is thus modelled by adding the variance of the probability distribution based on the observed data to the variance of the expected probability distribution of the replicating study based on its proposed sample size. This allows an estimate to be made of the probability that the P value will be less than or equal to 0.05 two sided (or any other specified value) in any replicating study. The same model can be used to estimate sample sizes when planning the required power of the initial study. If this requires doubling the variance, then this will require double the sample size estimated using current power calculations, suggesting that studies using current methods are underpowered. These considerations might be used to explain the replication crisis.

## 1. Introduction

It has been said that P values and confidence intervals have mystified generations of students (Spiegelhalter, 2019); and that learning from data is a bit of a mess (Ibid 2019). This is accompanied by persistent differences of opinion between Bayesians and Frequentists about how scientific data should be interpreted. A high-profile statement by the ASA has tried to clarify the position (Wassrerstein et al, 2016, Greenland et al, 2016). Nevertheless, it is still unclear how a P value is related to the probability of scientific replication as reflected by the ongoing replication crisis (Benjamin et al, 2018). In the Open Science Collaboration study, the average two-sided P value in 97 studies was 0.028 but only 36.1% (95% CI 26.6% to 46.2%) showed a two-sided P value of 0.05 or lower when each of the 97 studies was repeated (Open Science Collaboration, 2015). The estimation of probabilities and the replication of findings on which they are based in a single patient is important for clinical observations during medical diagnosis as well as for studies that test scientific hypotheses. The probability of replicating clinical observations reported by another clinician is a familiar daily experience for diagnosticians. This might provide an insight into the nature of replication in scientific studies. An understandable model of replication is also important for physicians who are trying to assess how well the result of a study elsewhere would be replicated in their own clinical setting.

2. Uniform priors, probabilities and likelihoods

Diagnosticians use an observed proportion based on clinical experience as a direct estimate of the probability of a diagnosis. They would also regard a continuous Gaussian distribution as directly estimated probabilities of possible parameters (e.g., true means) conditional on the information in the study's methods and data (e.g., an observed mean). However, statisticians would regard the Gaussian distribution as likelihoods of observing the methods and data (e.g., an observed mean and standard deviation) conditional on a range of possible parameters (e.g., possible true means). These view-points can be reconciled by assuming that the prior probability distributions of the parameters (e.g. possible true means) and data (e.g. possible observed means) are both uniform and on the same scale. This would be the case as the identical scales would be subsets of the universal set of all uniformly distributed numbers prior to any knowledge of the study's design being available and before a Bayesian prior probability can be estimated. Because of these uniform and equal prior probabilities of parameters and data, a Bayesian probability distribution of possible parameters conditional on personal experience is the same as the likelihood distribution of the same personal experience conditional on each possible parameter. According to this concept, multiplying a Bayesian distribution with a data distribution would be regarded as the first link in a potential chain of conditional independence.

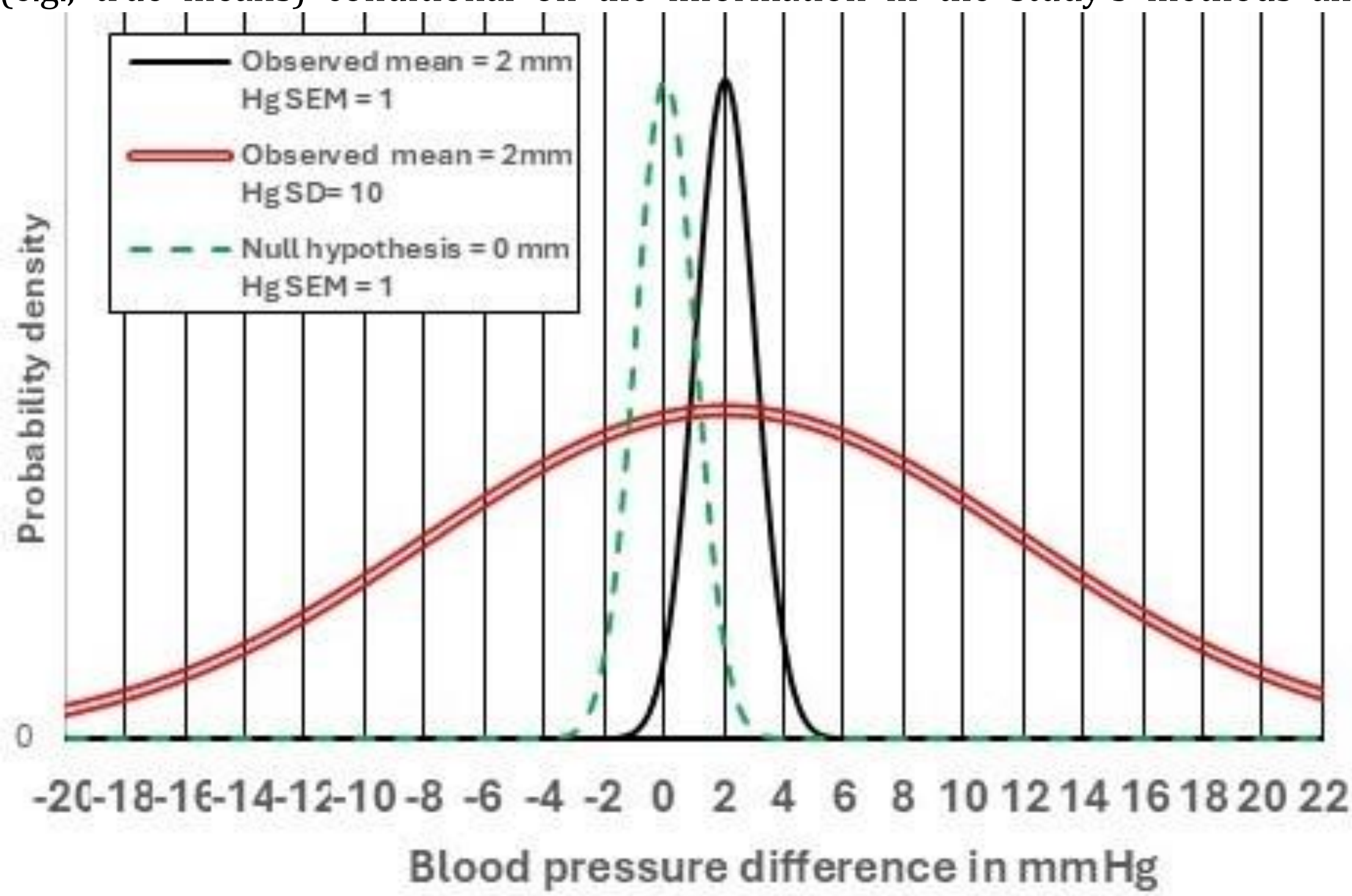

Fig. 1: The distributions of blood pressure differences in a cross-over RCT

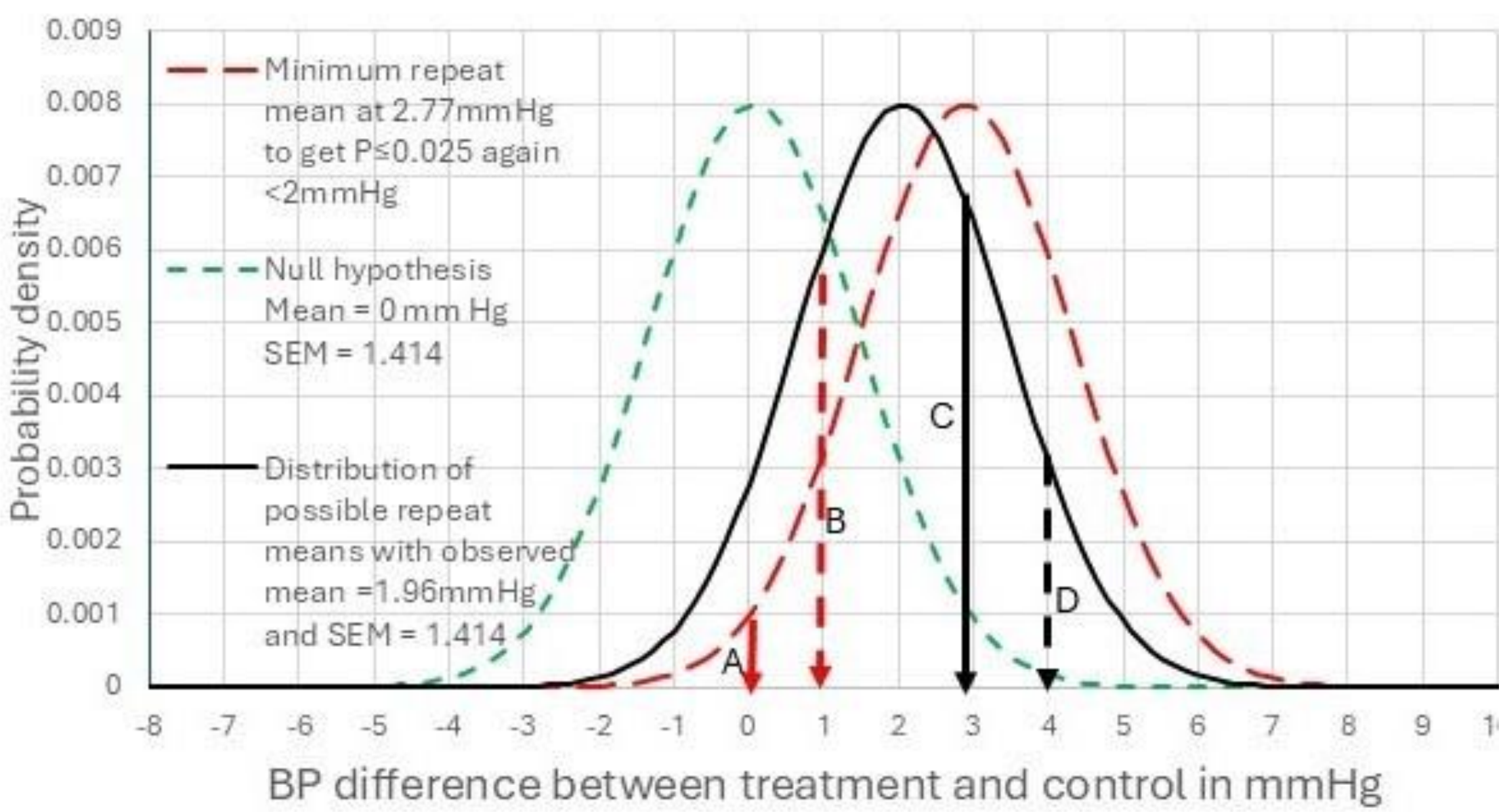

Fig. 2: The distribution of blood pressure differences when the results of one study are followed by the results of an identical study

## 3. A directly estimated probability of an outcome without Bayes rule

If 100 patients had been in a double-blind cross over randomized controlled trial and that 58 out of those 100 individuals had a BP that is 2 mm Hg higher on control than on treatment, then knowing only that an individual was one of those in the study, a diagnostician's probability conditional on the entry criterion of that individual patient having a BP difference greater than zero as a chosen range of interest would be 58/100 = 0.58. In Bayesian terms this would be a probability that is estimated directly without estimating prior probabilities, likelihood distributions and applying Bayes rule. If this BP difference of 2 mm Hg had been the result of an 'N of 1' study on a single patient that had been repeated randomly in that same person 100 times, then 0.58 would be the estimated frequency of getting a BP difference greater than zero if the BP was measured repeatedly in that individual.

### 3.1. Probabilities conditional on numerical results

If in a study on many subjects, the average BP difference between a pair of observations on treatment and control was 2 mm Hg and the standard deviation of the differences was 10 mm Hg, then in Figure 1, the area under the bell-shaped Gaussian distribution above 0 mm Hg to the right of the arrow (i.e. 0.2 standard deviations below the mean of 2 mm Hg) would contain 58% of the total area. From this again we see a probability of any randomly selected study individual will have a probability of 0.58 of being greater than zero. Therefore, fitting a Gaussian distribution to the data by calculating the standard deviation can be assumed to model the possible average results of continuing that study until there was an infinite number of subjects. This infinite number of possible results would create a distribution of an infinite number of possible continuous values as in Figure 1.

## 4. Distribution of possible means of different studies

As the standard deviation of the distribution was 10 mm Hg and the number of observations were 100, then the standard error of all the possible 'true' means is modelled by the expression $10/\sqrt{100} = 1$. Therefore, 95% of the area under the tall narrow curve in the centre of Figure 1 with a mean of 2 mm Hg and SEM of 1 mm Hg would fall between 0.04 mm Hg to 3.96 mm Hg. A distribution of means based on an infinite number of studies might be assumed to be a theoretical distribution of 'true' means. This implies that there is a theoretical probability of 0.95 that true mean lies between 0.04 mm Hg to 3.96 mm Hg. The one-sided probability would be 0.975 that the true mean is above 0.04 mm Hg. However, a scientist and diagnostician might wish to know the probability of a result being above 0 mm Hg as their range of interest for making clinical, research or policy decisions. This would be a probability that the true mean was 2 SEMs below 2 mm Hg after an infinite number of observations. This would be estimated from the Gaussian distribution as a theoretical probability of 0.9772. If their range of interest was a BP difference above 2 mm Hg then the probability would be 0.5 of course.

### 4.1. Different distributions of means

The distribution of possible means can represent 3 different situations: (A) The distribution of probabilities of all possible true means ('true' after an infinite number of observations) conditional on the observed mean and a standard error of the mean (SEM) of a limited measurement results (e.g., as represented by a distribution with a mean and mode at 2 mm Hg in Figure 1). The prior distribution of all possible hypothetical means conditional on the universal set of rational numbers can be assumed to be uniform (i.e., before the nature of any future study is known so that a Bayesian prior distribution cannot be estimated yet). Based on this uniform distribution conditional on the universal set of numbers and Bayes rule, the converse likelihood distribution of the observed mean conditional on all possible hypothetical means can be assumed to be the same as the corresponding probability distribution.

(B) The likelihood distribution of possible observed means conditional on a single hypothetical true mean by assuming a uniform prior probability of all hypothetical true means as in A so that the SEM and underlying mathematical model is the same as in A (e.g., the likelihood distribution with a mode at a null hypothesis of 0 mm Hg as in Figure 1).

(C) The likelihood distribution of a unique observed mean conditional on each of the possible hypothetical means, assuming that the SEM is the same as in A (e.g., as represented by a distribution with a mode at 2 mm Hg in Figure 1)

### 4.2. The null hypothesis

Based on situations A and B, then the tail area of the distribution in Figure 1 more extreme away than the observed mean (e.g., ≥ 2mm Hg in Figure 1) will be the same as the tail area of the distribution more extreme away than the null hypothesis (e.g. ≤ 0 mm Hg in Figure 1). This is because both distribution tails are the same number of SEMs away from the distribution mean. Therefore, the P value will be equal also to the probability of the true mean being more extreme away from the null hypothesis (e.g., less than 0 mm Hg) conditional on the observed mean (e.g., 2 mm Hg). So, the probability of the true mean being less extreme than the null hypothesis (e.g., >0 mm Hg) conditional on the observed mean will be equal to 1 minus the latter and 1 minus the P value. Thus, the likelihood of getting the observed mean difference of 2 mm Hg or something more extreme (i.e. over 2 mm Hg) conditional on the null hypothesis in Figure 1, is the P value of 0.0228. The probability of the true difference being of zero or lower is also 0.0228, and above zero it is 0.9772.

### 4.3. Uniform prior probabilities

It is assumed in this example that distribution (A) is Gaussian with a mean equal to the observed mean. It is also assumed that distribution (B) is a Gaussian distribution as in (A) and has the same SEM conditional on any hypothetical single mean (e.g. the null hypothesis). Therefore, when $X_i$ is any possible true mean and $X_{om}$ is the single observed mean, then $p(X_i|X_{om}) = p(X_{om}|X_i)$ and so by Bayes rule, $p(X_i) = p(X_{om})$ for any $X_i$ and therefore $p(X_i) = p(X_{i+1}) = p(X_{om})$ for any i. In other words, the latter are all the same so that the prior probability distributions of $X_i$ are uniform and equal to $p(X_{om})$.

## 5. Why the assumption of uniform priors on which P values are based seems reasonable

The assumption made historically to calculate P values is that the likelihood of selecting patients at random from a population of an assumed true value (e.g. when the null hypothesis is zero) is equal to the directly estimated probability distribution arising from the study. This assumes that scales used for the true and observed values are the same, unlike binomial distributions where this would be a special case [Llewelyn, 2019). The scale of continuous values used for the study are a subset of the universal set of all numbers, so it can be assumed that the prior probability of each possible value conditional on that universal set is the same or uniform for each of these true and observed values. This uniformity will apply to all studies using continuous numerical values and before any study design is considered.

### 5.1. Why Bayesian prior probabilities are not conditional on a universal set and not uniform

The Bayesian prior probability is different to the above uniform prior probability conditional on the universal set of all numbers. The Bayesian prior will be estimated not before but after designing the study and doing a thought experiment based on personal experience and reading the literature to estimate what the distribution of possible results might be in an actual study conditional on background knowledge. Any observed result will be independent of the prior estimated result and so that the former is not a subset of the latter, which is an essential condition for Bayes rule. In contrast both the numerical values of the prior and observed distributions will be subsets of the universal set of all numbers. Therefore, the Bayesian prior distribution can be regarded also as a posterior distribution formed by combining a uniform prior distribution conditional on the universal set of all numbers with an estimated likelihood distribution of the thought study result (or pilot study result) conditional on all possible true values. Each of these likelihoods is then multiplied by the likelihood of observing the actual study result conditional on all possible true results. These products are then normalized to give the Bayesian posterior probability of each possible true result conditional on the combined evidence of the result of the Bayesian thought experiment and the actual study result.

### 5.2. The baffled student

We might now explain to a student that a one-sided P value of 0.0228 is the same as the probability of the true study mean falling into a range beyond the null hypothesis conditional on the observed data, methods and mathematical models used. This also implies that there is a probability of 0.9772 that the true mean does fall into a range the same side as the observed mean. In some cases, the range of interest might be some other interval between two thresholds (e.g. a BP difference of between 0 mm Hg and 3.96 mm Hg or 1 mm Hg and 3 mm Hg). This happens of course when the investigator wishes to test the hypothesis that there is little difference between the outcome of treatment and control. In this case there would be a probability of 10.0228 - 0.0228 = 0.9544 that the true result mean would fall within the range of interest of 0 mm Hg to 3.96 mm Hg or a probability of 0.683 that true result mean will fall within the range of interest of 1 mm Hg and 3 mm Hg or a probability of 0.5 that it will fall with the range of interest over 2 mm Hg. It follows that there is a probability of 0.95 that the true result lies within the 95% confidence interval.

## 6. Replication

It is common experience that when a clinician is given a test result with a standard deviation, she or he would use this to predict a range of possible true results (e.g. with a probability of 0.95 within two standard deviations). If this range included the threshold for the normal range, it might cause hesitation and a decision to perform a second test and basing a decision on the average of the first and second results. The variance of this average result would be smaller and perhaps this time, two standard deviations might exclude the threshold for normal, thus increasing confidence in a decision. However, if the first test result was used to estimate the result a second test, then the numerical result might be the same, but the variance would be larger, perhaps doubled. In other words, when planning to 'replicate' a test result, the accuracy of the prediction is lower than when trying to predict a true result because the probability distribution for the second prediction is much wider than when trying to predict the 'true' result. Trying to predict a true result is analogous to shooting an arrow from a bobbing boat at a target on the shore. However, trying to predict a repeat result is analogous to shooting an arrow from a bobbing boat at a target on another bobbing boat. The same considerations apply when the result of one scientific study subject to variation is used to estimate the probability of another scientific study that attempts to replicate the first, which is also subject to variation.

### 6.1. The combined variance of two independent studies

To simplify matters, we will now use a slightly different example where the observed mean difference is 1.96 mm Hg instead of 2 mm Hg but the standard deviation is still 10 mm Hg and the SEM is still 1. However, if we postulate a distribution of differences for the result of the second study also with a SEM of 1 added to each point on the distribution of true mean values of the first study, then the added distribution of the second study results conditional on the information in the first study's methods section will have a variance of 1 + 1 = 2 and the SEM will be $\sqrt{2}$ = 1.414 mm Hg (see Figure 2). The latter also shows the indirect likelihood distribution conditional on the null hypothesis and the direct probability distribution conditional on the information in the first study's methods and results section (this new observed difference is used here to get the probability of a P ≤ 0.025 again when the original P value was also 0.025 as opposed to 0.0228 as in the initial example).

To get a combined result that provides a P value of exactly 0.025 again, the next replicating study's mean result needs to be 2.77 mm Hg that corresponds to arrow C in Figure 2. This is because the SEM of the distributions is 1.414 mm Hg and the result must be 1.96 SEMs away from the null hypothesis of zero. Therefore, the repeat result must be at least 1.414 mm Hg x 1.96 = 2.77 mm Hg away from zero, which is 2.77 -1.96 = 0.81 mm Hg away from an observed mean of 1.96 mm Hg. This corresponds to a Z score of -(0.81/1.414) = - 0.573. The area of the distribution with mean 1.96 mm Hg in Figure 2 to the right of arrow C is therefore 28.3% of its total area. In other words, when an observed difference from zero is 1.96 mm Hg, the standard deviation is 10 mm Hg and there are 100 observations, then the probability of getting a P-value of 0.025 or less when the study is repeated with the same number of observations is 0.283. The result of the calculation in a spreadsheet would be as follows:

=NORMSDIST((1.96)/(((10/100^0.5)^2)*2)^0.5+NORMSINV(0.025)) = 0.283 (Equation 1)

The probability of replication also applies to other interpretations of Figure 2. By assuming that the prior probability conditional on the set of all numbers is uniform (i.e. prior to knowing the nature of the study), then when P=0.025, the probability of the true value being a BP difference of >0 mm Hg is 0.975 (see the area to the right of arrow A in Figure 1 for the distribution with a mean of 1.96 mm Hg.. For a non-trivial difference of 1 mm Hg BP' we look at the area of the distribution with a mean of 1.96 mm Hg to the right of arrow B, where the probability of the true value being a BP difference of ≥ 1 mm Hg is 0.831 and P=0.169 (1-0.831). The probability of getting these same results again when the study is repeated remains 0.283.

If we move the arrow C to D (from a BP difference of 2.77 mm Hg to 3.77 mm Hg) then this BP difference of ≥ 3.77 mm Hg accounts for 10% of the results. They correspond to P≤ 0.003824 at the broken black arrow D for the distribution with a mean at the null hypothesis of zero. The probability of the true value being a BP difference of ≥ 0 mm Hg conditional on a result mean of 3.77 mm Hg is 0.996176 (1-0.003824). This is represented in Figure 2 by moving the distribution from a mean of 2.77 mm Hg (the big black arrow) so that the mean is 3.77 mm Hg (the small broken arrow D). However, the probability of the true value being a BP difference of ≥ 1 mm Hg conditional on an observed mean of 3.77 mm Hg is 0.975. There is a probability of 0.100 that this will also be the case if the study is repeated (corresponding to the area of the distribution with a mean of 1.96 mm Hg to the right of arrow D):

=NORMSDIST((1.96)/(((10/100^0.5)^2)*2)^0.5+NORMSINV(0.003824)) = 0.100 (Equation 2)

6.2. General considerations

The above example is a special case of course, where the standard deviation is 10 mm Hg, the number of observations 100 and there is a requirement for the P value resulting from the repeat experiment with the same number of measurements is ≤ 0.025. However, if the above result had been based on 409 observations instead of 100, the probability of replication with a P value of ≤ 0.025 again would have been:

=NORMSDIST((1.96)/(((10/409^0.5)^2)*2)^0.5+NORMSINV(0.025)) = 0.8 (Equation 3)

If we had been planning the above study and had estimated that the difference between the observed mean difference and zero was 1.96 mm Hg and that the standard deviation was 10 mm Hg then by inverting Equation 3, the estimated number of observations required to get $P \leq 0.025$ in the proposed study based on 2 variances (see '/2' in the expression below) would round up to 409:

=(10/(((1.96/(NORMSINV(0.8)-NORMSINV(0.025)))^2)/2)^0.5)^2 = 408.6 (Equation 4)

This estimate is based on a double variance whereas the current approach to estimating sample sizes is based only on one variance, which is the likelihood distribution of possible results based on a single estimated difference of the mean and standard deviation. As a result, the conventional estimate for the number of observations required based on one variance (see '/1' in the expression below) to get $P \leq 0.025$ in the proposed study would round up to 205:

=(10/(((1.96/(NORMSINV(0.8)-NORMSINV(0.025)))^2)/1)^0.5)^2 = 204.3 (Equation 5)

If we used this number in planning the study, then according to the above argument, the probability of a result with $P \leq 0.025$ one sided again is only about 0.5, which agrees with the Goodman's estimate (Goodman 1992):

=NORMSDIST((1.96)/(((10/205^0.5)^2)*2)^0.5+NORMSINV(0.025)) = 0.510 (Equation 6)

If we wished to get $P \leq 0.025$ in a second statistically independent study with a probability of 0.8 (i.e. a power of 80%) to replicate the first proposed study, then we would need to triple the variance (see '/3' in the expression below). We would then require about 613 observations:

=(10/(((1.96/(NORMSINV(0.8)-NORMSINV(0.025)))^2)/3)^0.5)^2 = 612.9 (Equation 7)

This means that the probability of getting $P \leq 0.025$ in the planned first study would be very high at 0.929 based on 613 observations:

=NORMSDIST((1.96)/(((10/613^0.5)^2)*2)^0.5+NORMSINV(0.025)) = 0.929 (Equation 8)

However, the probability of getting $P \leq 0.025$ in the planned second replicating study would be 0.8 as expected:

=NORMSDIST((1.96)/(((10/613^0.5)^2)*3)^0.5+NORMSINV(0.025)) = 0.800 (Equation 9)

If the result of the first study becomes known and happens to have an observed mean difference of 1.96 mm Hg and a standard deviation of 10 mm Hg (as estimated at the planning stage), then when 409 subjects have been recruited so far, the estimated probability of replication in the second replicating study would be based on 2 variances ('*2/') in equation 10 below instead of 3 variances ('*3/') in equation 9 above. The probability of replication is then also 0.8:

=NORMSDIST((1.96)/(((10/409^0.5)^2)*2)^0.5+NORMSINV(0.025)) = 0.8 Equation 10

Instead of assuming that the number of observations is the same in the completed first study and the second replicating study (i.e. 409) we can estimate the probability of replication if we were to make a different number of observations in the second replicating study (e.g. 800 as shown in equation 11 where the number of observations for the first and second study are specified separately:

=NORMSDIST((1.96)/(((10/409ˆ0.5)ˆ2)+((10/800ˆ0.5)ˆ2))ˆ0.5+ NORMSINV(0.025))= 0.897 (Equation 11)

According to the model and assumptions put forward here, the currently conventional calculated number 205 of observations required to get $P \leq 0.025$ in the first planned study appears to have a power of only 50%. To have a power of 80%, this conventional number needs approximately to be doubled to 409. Furthermore, to have a power of 80% of getting $P \leq 0.025$ in the planned second replicating study using the same number of observations, the number needs to be tripled to 613.

We can go further and expect the second study to have an infinite number or perhaps millions of observations, which would be an ideal test of replication. Thus, if the first study based on 205 observations turned out to have a standard deviation of 10 mm Hg and a mean of 1.96 mm Hg, but a second study had a sample size of 10 million ('10000000' in Equation 12 below) the probability of replication with a P value of 0.05 or less two sided again would be:

=NORMSDIST((1.96)/(((10/205ˆ0.5)ˆ2)+((10/10000000ˆ0.5)ˆ2))ˆ0.5 +NORMSINV(0.025)) = 0.801 (Equation 12)

If we used 205 observations from the currently conventional calculation to estimate the probability of the first study giving a P value of 0.05 or less, it would be about 0.5 as shown in Equation 6. The probability of getting a P value of 0.05 or less in both studies of the same size (without knowing the result of the first study) would be based on 3 variances ('*3/' in Equation 13) This probability would be about 0.367: =NORMSDIST((1.96)/(((10/205^0.5)^2)*3)^0.5+NORMSINV(0.025)) = 0.367 (Equation 13)

7. The Open Science Collaboration study

In the Open Science Collaboration (2015) study, the average P value from the original 97 articles was 0.028 two sided but only 35/97 = 36.1% (95% CI 26.6% to 46.2%) showed a two-sided P value of 0.05 or lower when repeated. If we assume that all the studies had been planned by using conventional calculations to provide an expected power of 80%., then according to the reasoning in this paper and Equation 6, the true power would be 51%. (i.e. with a probability of 0.51 of getting a P value of 0.05 or less). It follows that the probability of getting a P value of 0.05 or less on both occasions would be about 0.367 as shown in Equation 13, which is what Open Science Collaboration discovered at 36.1%.

8. The size of the second replicating study

In Equation 6, the calculated probability of replicating an observed result with a P value of at least 0.05 two sided when the second study was the same conventional sample size of 205 as the first study was 0.5, as suggested by Goodman (1992). However, Senn (2002) pointed out that to specify that the second replicating study should be the same size as the first study was unnecessary and a larger number could be considered for the second replicating study. If we postulate that the second replicating study was infinitely large, then the probability (e.g., of 0.8) of the second study also having a P value of no higher than 0.05 two sided would be equal to the estimated power of the first study (e.g., 80%) as shown in Equation 12. This would be difficult to demonstrate empirically with very large second studies, but the Open Science Collaboration (2015) study result is consistent with what we should expect from the reasoning model in this paper, so its result supports it.

If the second replicating Open Science Collaboration (2015) study had been assumed to infinitely large, then the probability of replication based on the same sample sizes should be 0.51 and that the frequency of replication would be expected to be about 51%. If the sample size assumed to calculated in a conventional way were doubled, then the probability of a P value being 0.05 or less in the first study would have been 0.8 and the frequency of replication with a P value of 0.05 or less in a theoretical replicating study with an infinitely large number of subjects would be 80% also.

9. The relationship between P values and confidence intervals

When P = 0.025 and the null hypothesis is zero, then by assuming uniform prior probabilities of numerical results conditional on the universal set of all numbers before the nature of the study is known, the 95% confidence interval will also just exclude zero. Therefore, the probabilities in relation to P values in Equations 1 to 13 also applies to calculated 95% confidence interval in subsequent studies also excluding zero. for example, to expect the probability of 0.975 for the true mean of a second study to exceed zero, the 100 observations made already in Figure 1 should be increased to 205. This is because if $s$ is the standard deviation and $n$ the number of observations, the SEM of the first study is $s/\sqrt{n}$, then the variance is $s^2/n$. The expected 'double variance' by also expecting the second study to exclude zero is $2s^2/n$. If we double the number of observations made already from n to 2n, the expected combined variance for the second study will be $2s^2/2n = s^2/n$, so that the SEM is $s/\sqrt{n}$, so the confidence interval will be the same as in the first study.

10. Summarizing the relationship between sample size and probability of replication

Current sample size calculations appear to result in under-powered studies and in general sample sizes should be doubled to achieve a desired power (e.g of 80%). Replication probabilities based on second studies should be estimated based on an infinite sample size 'in the long run' instead of the same small unreliable 'short term' size. For example, when planning a study with an estimated mean of 1.96 and a standard deviation of 10:

1. A sample of 205 based on a conventional 80% power calculation only gives a0.5 probability of getting a P value of 0.05 in the first study and if it is repeated in a second study with the same number there is only a probability of 0.367 of getting a P value of 0.05 (as demonstrated by the Open Science Study). However, if we were to use an infinite sample size in the second study, then there is a probability of 0.5 of getting a P value of 0.05 or less and a probability of 0.5 that the calculated 95% confidence interval will exclude zero.
2. A sample of 409 gives a 0.8 probability of getting a P value of 0.05 or less in the first study. If the study is repeated with the same sample size, there is only a probability of 0.5 of getting a P value of 0.05 or less in the second study. However, if we were to use an infinite sample size in the second study, then there is a probability of 0.8 of getting a P value of 0.05 or less and a 90% confidence interval that excludes zero.

3. A sample of 613 gives a 0.929 probability of getting a P value of 0.05 or less in the first study and if it is repeated in a second study with the same sample size, there is a probability of 0.8 of getting a P value of 0.05 or less in the second study. However, if we were to use an infinite sample size in the second study, then there is a probability of 0.929 of getting a P value of 0.05 or less and a 95% confidence interval that excludes zero.

## 11. Investigating whether the apparent true result is due to bias etc.

The probability (e.g., 0.975) of a true mean being within a range of interest only suggests that the result of the study is promising. This could be interpreted by stating that the probability of it being the true result is up to 0.975 (i.e., 0 to 0.975). This result may not be due to the phenomenon of interest (e.g. the treatment being more effective than placebo). It could be due to bias or dishonesty or cherry picking or data dredging etc. This list of possible causes of the study result can be investigated by looking for items of evidence that hopefully make each of the invalid causes of the result unlikely, leaving a genuine treatment effect as the probable explanation. Mayo (2018) calls this process severe testing. This is analogous to a physician working logically through a list of differential diagnoses.

The probability theory for investigating rival diagnoses or scientific hypotheses can be modelled with a derivation of Bayes rule with a dependence assumption (Llewelyn et al, 2014). This derivation was formulated to model the differential diagnostic process as performed by physicians. It can use ratios of probabilities between pairs of rival diagnoses (or rival scientific hypotheses) in addition to ratios of likelihoods between two diagnoses. However, it can not only provide a probability of a diagnosis being correct, but the probability of a hypothesis or some other explanation not yet considered to be true (so called ’abductive’ reasoning). Failure to apply this ’severe’ testing’ by elimination as also suggested by Mayo (2018) to possible causes of bias etc. before publication may also partly explain the low frequency of replication currently being experienced. The validity of these reasoning models could be assessed empirically with the outcomes of preregistered clinical trials and other studies.

## 12. Testing scientific hypotheses

We might not only be interested in the effectiveness of a treatment (e.g. of an angiotensin receptor blocker or ‘ARB’ being better than placebo) but also in the theories of the underlying mechanisms. For example, if we had already found evidence to support the efficacy of an ARB in preventing nephropathy, we might also wish to do another study to compare the efficacy of an ARB in preventing nephropathy with another treatment that does not theoretically block the angiotensin receptor resulting in reduced protein leakage, but which lowers the BP to the same extent. If the ARB was shown to be more effective than a non-ARB in this situation, then this would support the hypothesis that this was due to the ARB reducing renal protein leakage in addition to reducing blood systemic pressure. The result would not confirm the angiotensin receptor theory because there might be another explanation that had not been considered that is also compatible with the findings. This is in keeping with Karl Popper’s assertion that hypotheses cannot be proven but only that rival hypotheses can be refuted (or at least shown to be less probable).

## 13. Conclusion

The principle of estimating probabilities directly from observed frequencies during diagnostic reasoning and seeking justification for this offers further insight into the relationship between probability of replication and P values. This results in a proposed model based on adding the variances of an original study and a proposed second study that is hoped will replicate the result of the first study. The results of these calculations can explain the low frequency of replication currently being experienced and how it can be improved by increasing sample sizes at the planning phase. The validity of these multiple variance models could be assessed more formally via the outcomes of preregistered clinical trials and other studies. A reliable model of replication is important for readers of clinical trial reports such as physicians who are trying to assess how well a study would be replicated in their clinical setting. These replication models might also be helpful in teaching and understanding statistics.

## 14. Competing interests

No competing interest is declared.

## 15. Author contributions statement

HL conceived the work, wrote and reviewed the manuscript.

## 16. Acknowledgments

The author thanks members of the discussion groups DataMethods for their valuable suggestions.

References

Benjamin, D. J., Berger, J. O., Johannesson, M., Nosek, B. A., Wagenmakers, E. J., Berk, R., ... Johnson, V. E. Redefine statistical significance. Nature Human Behaviour, 2018. 2(1), 6-10. https://doi.org/10.1038/s41562-017-0189-z

Goodman, S. A comment on replication, P-values and evidence. Statistics in Medicine 1992, 11(7): 875–879. pmid:1604067 Greenland S, Senn SJ, Rothman KJ, Carlin JB,

Poole C, Goodman SN, Altman DG. Statistical Tests, P-values, Confidence Intervals, and Power: A Guide to Misinterpretations. Online Supplement to the ASA Statement on Statistical Significance and P-values. The American Statistician, Volume 70, 2016 - Issue 2, p 129 – 133.

Llewelyn H. Replacing P-values with frequentist posterior probabilities of replication—when possible parameter values must have uniform marginal prior probabilities. PLoS ONE (2019) 14(2): e0212302.

Llewelyn H, Ang AH, Lewis K, Abdullah A. The Oxford Handbook of Clinical Diagnosis, 3rd edition. Oxford University Press, Oxford. 2014 p638 - 642.

Mayo D. Statistical Inference as Severe Testing: How to Get Beyond the Statistics Wars. Cambridge: Cambridge University Press, 2018.

Open Science Collaboration. Estimating the reproducibility of psychological science. Science. 2015; 349 (6251):aac4716.0

Senn, S. A comment on replication, p-values and evidence by S.N. Goodman. Statist. Med. 2002; 21:2437–2444 (DOI: 10.1002/sim.1072)

Spiegelhalter D. The art of statistics. Learning from data. Penguin Random House, 2019. p. 241.

Spiegelhalter D. The art of statistics. Learning from data. Penguin Random House, 2019 p 305

Wasserstein RL, Lazar NA. The ASA's Statement on p-Values: Context, Process, and Purpose. The American Statistician, Volume 70, 2016 - Issue 2, p 129 – 133.